\documentclass[10pt]{article} 
\usepackage{spie}
\usepackage{latexsym}
\usepackage{times,mathptm} 
\usepackage{graphicx} 
\usepackage{multicol}
\usepackage{rotating} 
\usepackage[verbose]{wrapfig}
\usepackage{epsfig}
\newlength{\dinwidth}
\newlength{\dinmargin}
\newcommand{\bc}{\begin{center}}
\newcommand{\ec}{\end{center}}
\newcommand{\be}{\begin{equation}}
\newcommand{\ee}{\end{equation}}
\newcommand{\bi}{\begin{itemize}}
\newcommand{\ei}{\end{itemize}}
\newcommand{\bt}{\begin{table}}
\newcommand{\enta}{\end{table}}
\newcommand{\g}{$\gamma$}

\newcommand{\cs}{$^{137}$Cs\ }

\newcommand{\na}{$^{22}$Na\ }

\newcommand{\yt}{$^{88}$Y\ }

\newcommand{\beq}{\begin{equation}}
\newcommand{\eeq}{\end{equation}}
\newcommand{\bfg}{\begin{figure}}
\newcommand{\efg}{\end{figure}}
\newcommand{\keV}{\mbox{ke\hspace{-0.1em}V}}
\newcommand{\MeV}{\mbox{Me\hspace{-0.1em}V}}

\newcommand{\phibar}{\ensuremath{\bar{\varphi}}}
\newcommand{\phigeo}{\ensuremath{\varphi^\triangleleft}}
\renewcommand{\deg}{\ensuremath{^\circ}}

\newcommand{\rsi}{\ensuremath{\vec{r}_{\star i}}}

\graphicspath{{figures/}}               

\begin{document}
\title{Spectroscopy and Imaging Performance of the Liquid Xenon Gamma-Ray
Imaging Telescope (LXeGRIT)}  
%
\author{
        E.~Aprile$^a$, A.~Curioni$^a$, V.~Egorov$^a$,
	K.-L.~Giboni$^a$, U.G.~Oberlack$^a$, \\
        S.~Ventura$^{a,b}$, T.~Doke$^c$, J.~Kikuchi$^c$,
        K.~Takizawa$^c$, \\
	E.L.~Chupp$^d$, P.P.~Dunphy$^d$ \\ 	
        \skiplinehalf
        $^a$Columbia Astrophysics Laboratory, Columbia University \\
	$^b$INFN--Padova, Italy \\
	$^c$Waseda University, Japan \\
	$^d$University of New Hampshire, USA \\
}
\authorinfo{Send correspondence to: E.~Aprile, Columbia University, 
Astrophysics Laboratory, 550 West 120th Street, New York, NY 10027 \\
E-mail: age@astro.columbia.edu \hfill \\
LXeGRIT Web page: \texttt{http://www.astro.columbia.edu/$\sim$lxe/lxegrit/}
}
\pagestyle{plain}    
\maketitle

\begin{abstract}
   
LXeGRIT is a balloon-borne Compton telescope based on a liquid xenon time
projection chamber (LXeTPC) for imaging 
cosmic \g--rays in the energy band of 0.2- 20 \MeV.  The detector, with
400~cm$^2$ 
area and 7 cm drift gap, is filled with high purity LXe. Both 
ionization  and scintillation light signals are detected to measure the energy
deposits  and the three spatial coordinates of individual \g --ray interactions
within the sensitive volume.
The TPC has been characterized with repeated measurements of its spectral and
Compton imaging response to \g --rays from radioactive sources such as \na,
\cs, \yt and Am-Be.
The detector shows a linear response to \g --rays in the energy range 511~\keV --4.4~\MeV, 
with an energy resolution (FWHM) of $ \Delta E/E=8.8\% \: \sqrt{1\MeV /E}$. 
Compton imaging of \yt \ \g --ray events with two detected interactions is
consistent with an angular resolution of $\sim$ 3 degrees
(RMS) at 1.8~\MeV. 

\end{abstract}

\keywords{gamma-rays, instrumentation, imaging, telescope, balloon missions, 
high energy astrophysics}

\section{Introduction}

Progress in MeV \g --ray astrophysics has lagged far behind that in the X--ray
band and at high energies, because of the difficulty of imaging \MeV\ photons
combined with the high background and the low cosmic source fluxes. Results from 
CGRO--COMPTEL \cite{VSchon:93},the only Compton Telescope in space to-date,
have shown both the promise and the challenges of this field. To explore the
rich scientific potential of this energy band, new instruments are needed 
which  combine substantially higher detection efficiency and background
rejection capabilities, within a large field-of-view. \\
The development of a liquid xenon time projection chamber (LXeTPC) at
Columbia was initiated to demonstrate the feasibility of the technique and to
study its capability as efficient  Compton telescope for \MeV\ \g --ray
astrophysics.   
The combination of calorimetry and 3D event imaging in a TPC is especially
powerful for reconstructing the multiple Compton histories of \MeV\ \g --rays,
and thus the incoming direction and energy on an event-by-event basis. Equally
important, the imaging capability proves to be a powerful tool in rejecting
background. 
After the characterization in the laboratory, the LXeTPC has been turned
into  a balloon-borne instrument (LXeGRIT) to test its performance in the near
space environment. A description of the instrument and its performance during
its May 1999 balloon flight in New Mexico are reported elsewhere in these
proceedings \cite{EAprile:SPIE2000}.   
Here we present some of the results on the detector spectral and imaging
performance, obtained during the pre-flight calibration experiments  with
radioactive sources of gamma-rays at energies between 511 \keV\ and 4.4 \MeV.  

\section{The LXeTPC as $\gamma$-ray spectrometer and Compton imager}
        
\begin{figure}[p]
\centering
\psfig{file=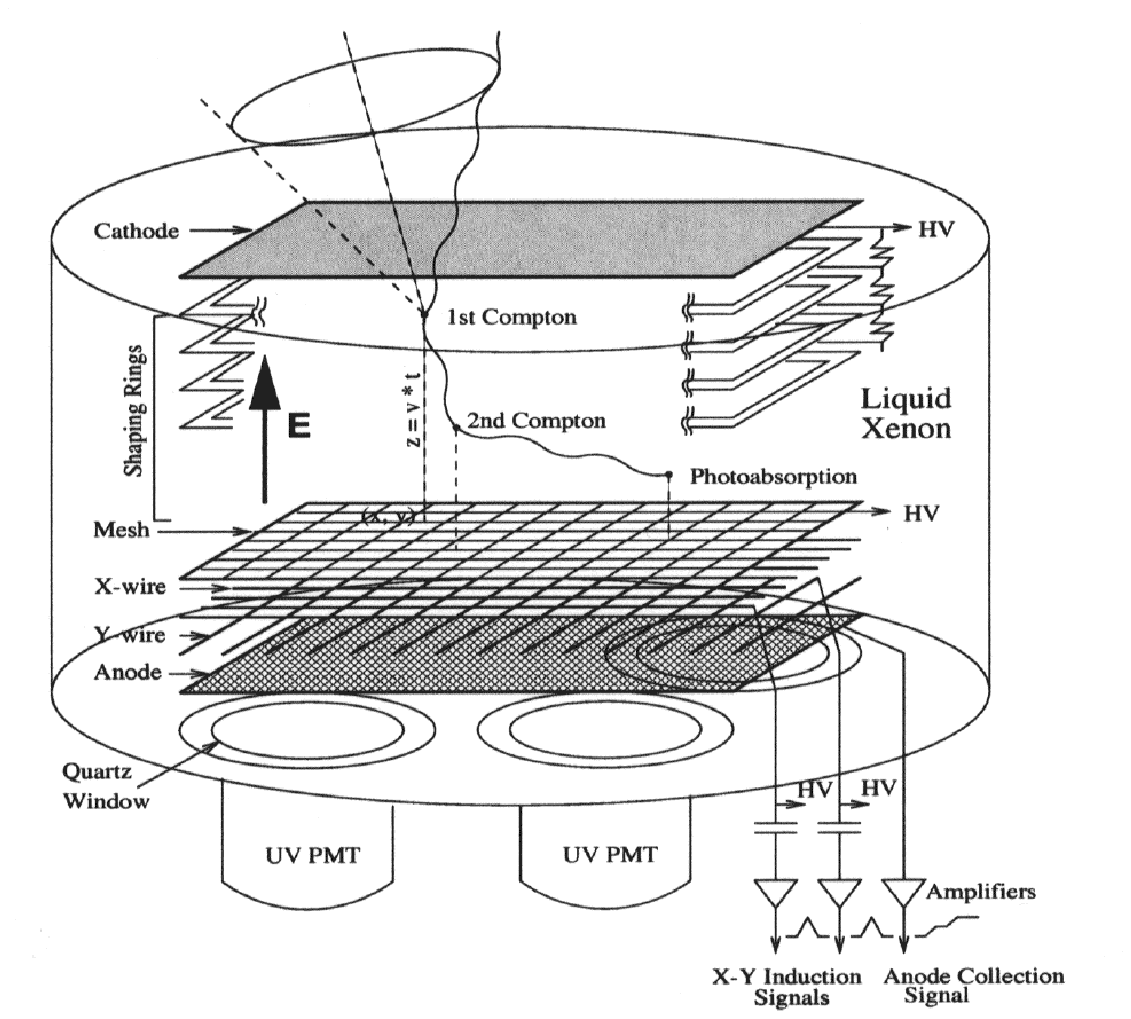,width=0.75\linewidth,clip=}
\caption{\label{f:TPC:schem}LXeTPC schematic.}
\bigskip
\psfig{file=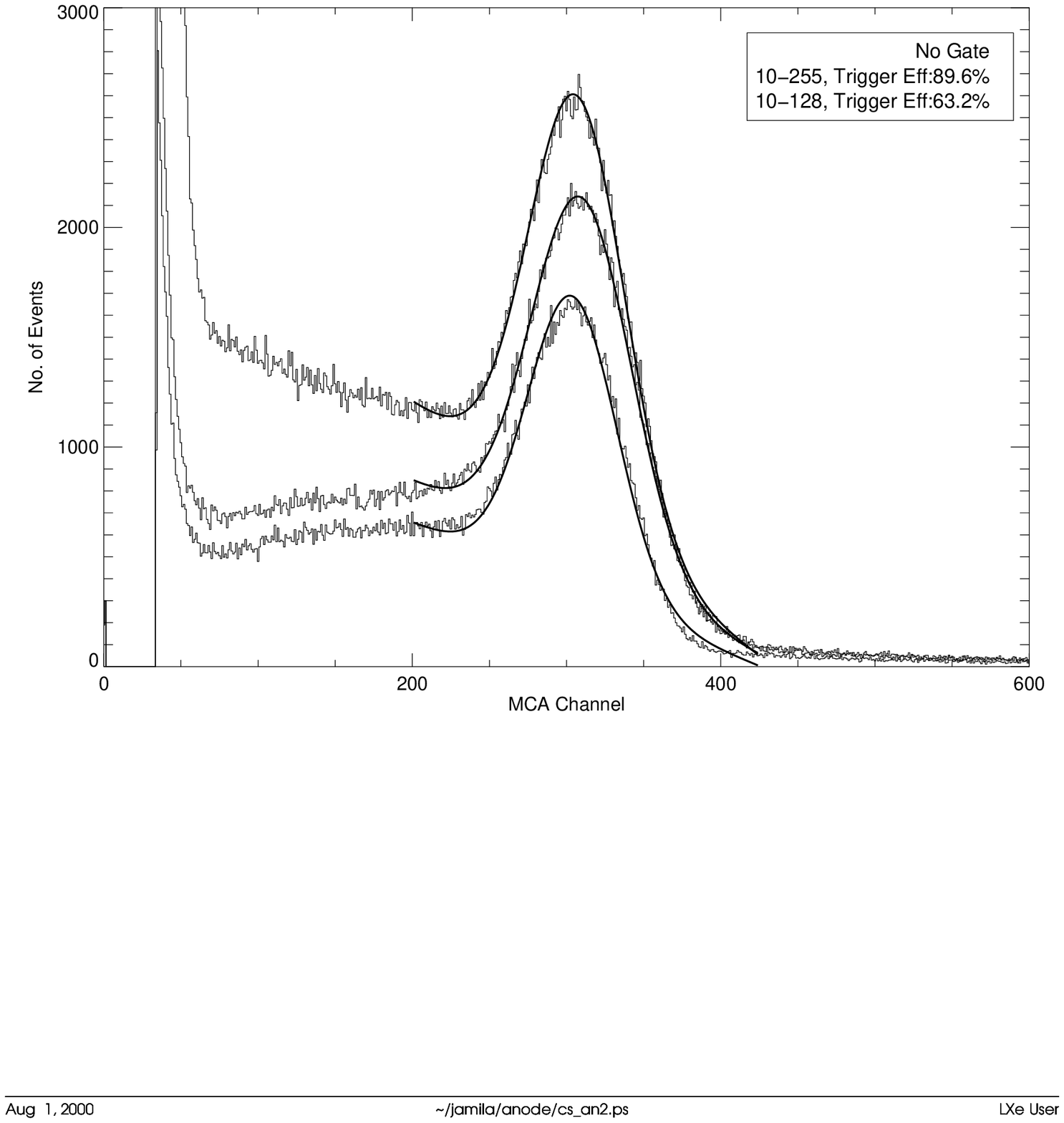,bbllx=50,bblly=360,bburx=600,bbury=760,%
       width=.6\linewidth,clip=}
\caption{\label{f:light_1} \cs 662~\keV \ energy spectra for all anode signals
({\it upper curve}) and for anode signals with a light trigger ({\it lower
curves}) for two discriminator windows.}
\end{figure}

A schematic of the LXeTPC and its principle of operation is shown in
Fig.~\ref{f:TPC:schem}. The TPC is assembled in a cylindrical vessel of 10~l
volume, filled with high purity LXe at a temperature of $\sim -100\deg$~C. The
sensitive area is $20 \times 20$~cm$^2$ and the maximum drift length is 7 cm. The
detector  operates over a wide energy range from $\sim$~200~\keV\ to 20~\MeV.   
\subsection {Light signals detection and processing}

Both ionization and scintillation light signals, produced by a
charged particle in the liquid, are detected to measure its energy and 3D
position. The fast ($<5$~ns) Xe light (175~nm), detected by four UV
sensitive photomultiplier tubes (PMTs)
coupled to the detector vessel by quartz windows, provides the event
trigger.  
The signals from the PMTs are passed trough a set of window discriminators: the
lower threshold of the discriminators avoids triggering the readout on noise
pulses, while the upper threshold reduces the number of high-energy
charged particle events. The choice of the upper and lower threshold, together
with the PMTs 
high-voltage, impacts significantly the detector's trigger efficiency.\\
The light level seen by the PMTs is extremely
low. The total number of photons produced in LXe by a charged particle is
reduced by the quenching of the recombination luminescence due to the
applied  electric field. More photons are lost due to
geometrical effects (solid angle), low reflectivity of the chamber materials,
optical transmission of the wires and anode meshes and transmittance by the
quartz windows.
The efficiency of each PMT was estimated by measuring the
count rate of the triggering  events relative to the total
count rate. This was done by comparing the Multi Channel Analyser (MCA) spectrum
of the charge signals from the anode 
directly above the PMT, gated by the PMT light trigger and ungated.
The ratio of events with a trigger and of all events, under
the full energy peak, was taken as the trigger efficiency 
for the particular $\gamma$-ray energy line. 
Three spectra for a collimated \cs\ source, placed above the center of one
anode, are shown in Fig.~\ref{f:light_1}. The voltage on the corresponding PMT 
was 1800 V. The MCA spectra were taken with 10~$\mu$s 
shaping time for the anode shaping amplifier. Given the 35~$\mu$s maximum
drift time, 
\g-rays interacting two or more times in locations more than 10~$\mu$s apart in
drift time contribute to the background below the line. 
From a fit of the lines, the trigger efficiency at 662~\keV\ is 
89.6~$\%$ for the discriminator window of 10--255. By lowering the
upper threshold from 255 to 128, the efficiency drops to 63~$\%$.  

\subsection {Charge signals detection and processing}

The drift of free ionization electrons in the LXeTPC uniform  electric field,
typically 1~kV/cm, induces charge signals on a pair of orthogonal planes of
parallel wires with a 3~mm pitch, before collection on four independent
anodes (see Fig.~\ref{f:TPC:schem}). Each of the 62 X-wires and 62 Y-wires and each anode is amplified and
digitized at a sampling rate of 5~MHz, to record the pulse shape. The X-Y
coordinate information is obtained from the pattern of hits on the wires, while
the energy is obtained from the amplitude of the anode signals. The
Z-coordinate is determined from the drift time measurement, referred to the
light trigger.
For a review of the TPC signal characteristics and their electronics readout we
refer to Aprile et al. 1998 \cite{EAprile:98:electronics}. The system is such
that the equivalent 
noise charge on the wires is typically less than $\sim$~400~e$^-$~RMS, while the
noise on the anodes, of higher capacitance, is $\sim$~800~e$^-$~RMS. With these
noise conditions, the TPC can well detect the multiple interactions of \MeV \
\g--rays, with a minimum energy deposit of $\sim$150~\keV. \\
Fig.~\ref{f:sig_rec} shows the TPC display of a \yt\ 1.8 \MeV\
\g--ray event.  
The digitized signals on all wires and active anodes are shown as a function of
drift time. The incoming photon makes two Compton scatterings befor being
photoabsorbed. In this ``3--steps'' event, in fact, the sum of the three anode
pulse heights gives 1.8~\MeV. The corresponding location of the interactions are
clearly seen on the X-Y wires.   
An efficient and fast signal recognition algorithm has been developed to sort
the different event topologies and to obtain the
energy and X-Y-Z coordinates from the digitized signals of each interaction recorded in the sensitive
volume \cite{oberlack_1}. With this information, and with the gain and energy
calibration curves, spectral and imaging analysis of gamma-ray sources is then
carried out.

\bfg[htb]
\centering
\psfig{file=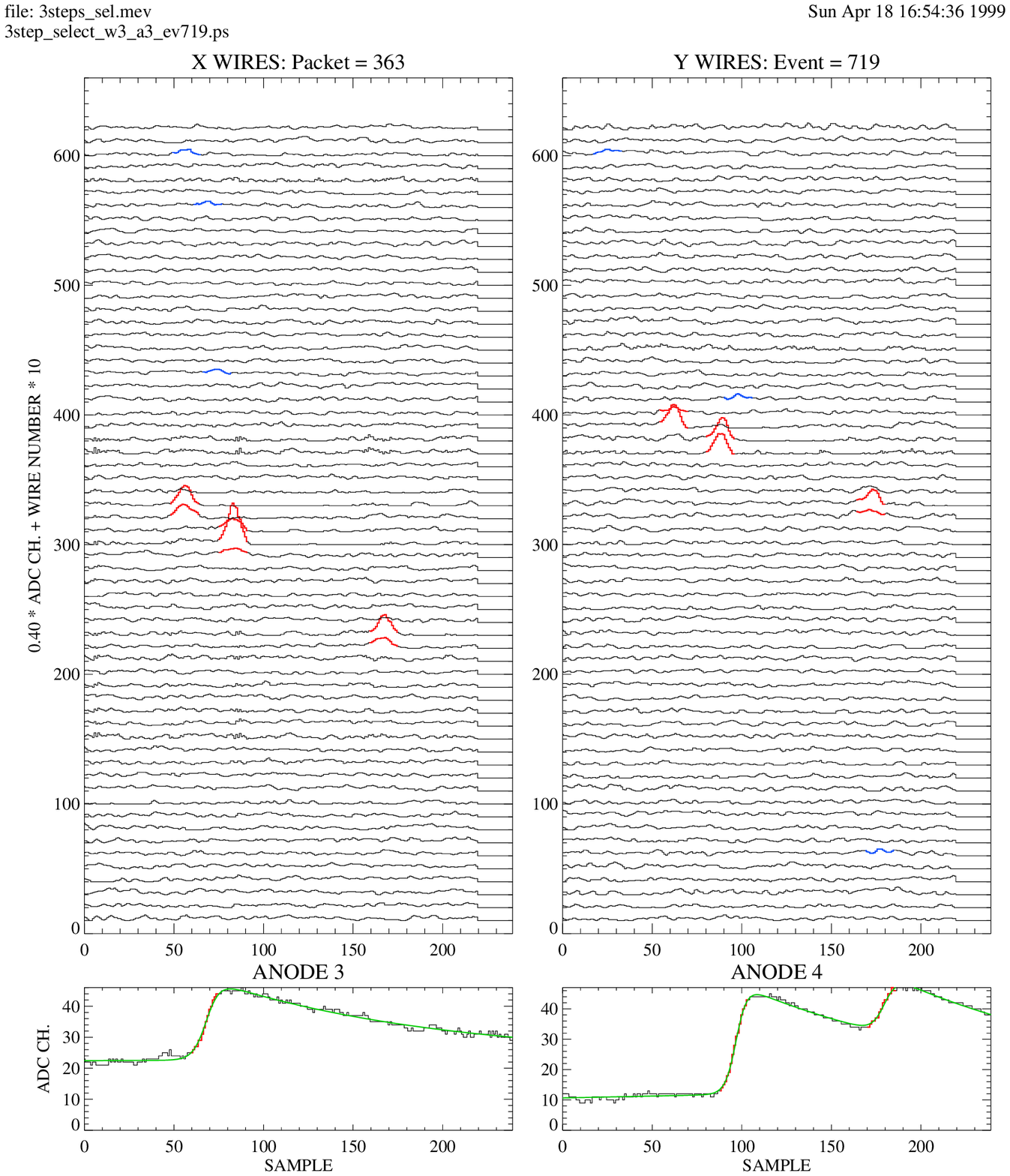,bbllx=62,bblly=143,bburx=563,bbury=726,%
       width=.7\linewidth,clip=}
\caption{\label{f:sig_rec} LXeGRIT on-line display of a 1.8~\MeV\ \g-ray
event with multiple Compton interactions.}  
\efg

\subsection {\label{s:ecal} Energy Resolution and Calibration}

\begin{figure}[p]
\centering
\epsfig{file=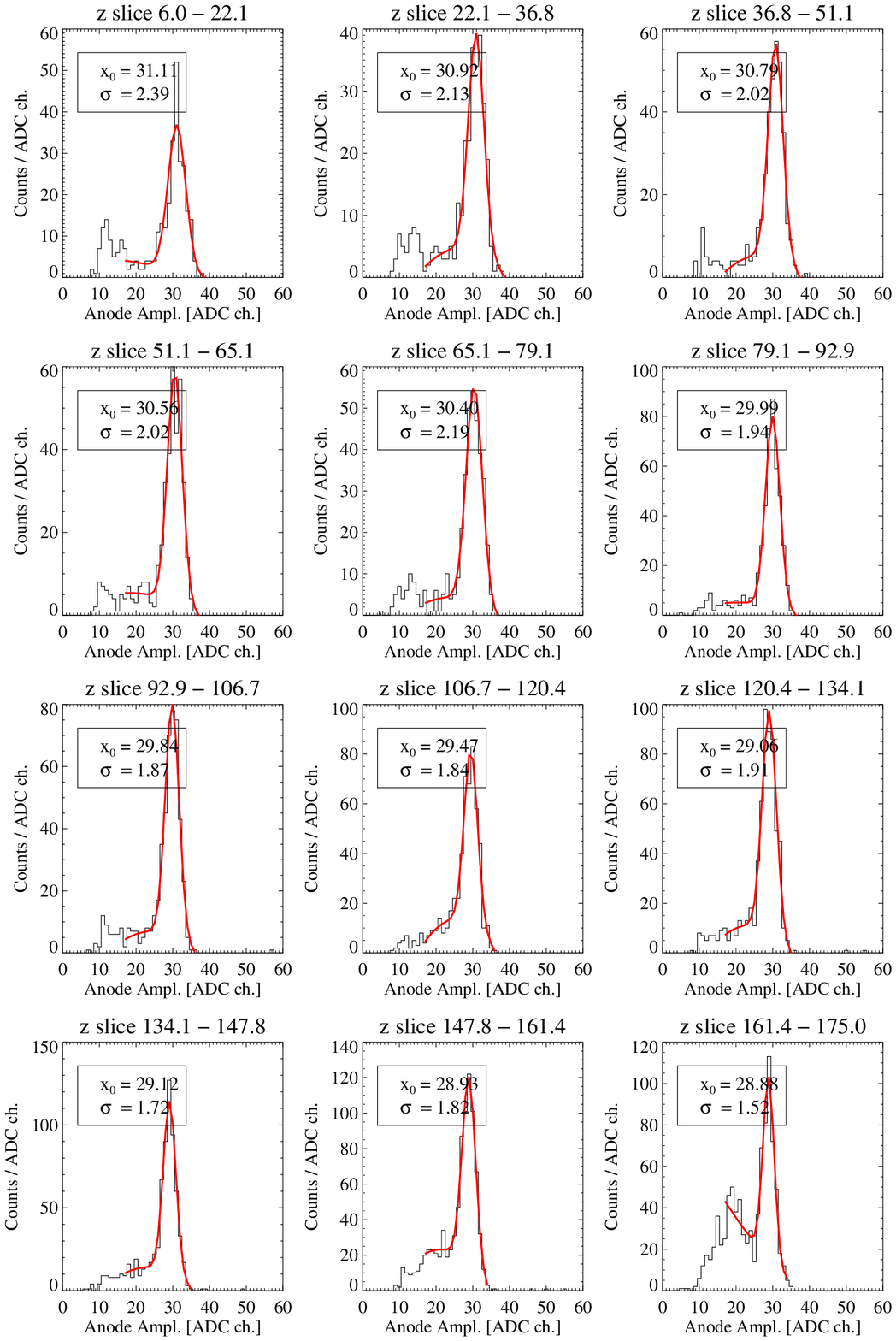,bbllx=50,bblly=75,bburx=580,bbury=750,width=.6\linewidth,clip=}
\caption{\label{f:cs_slic} \cs (collimated source) energy spectra for different
drift times: each ``slice'' of the fiducial volume is 6~mm thick. Going from top
to bottom the drift time increases.}  
\bigskip
\epsfig{file=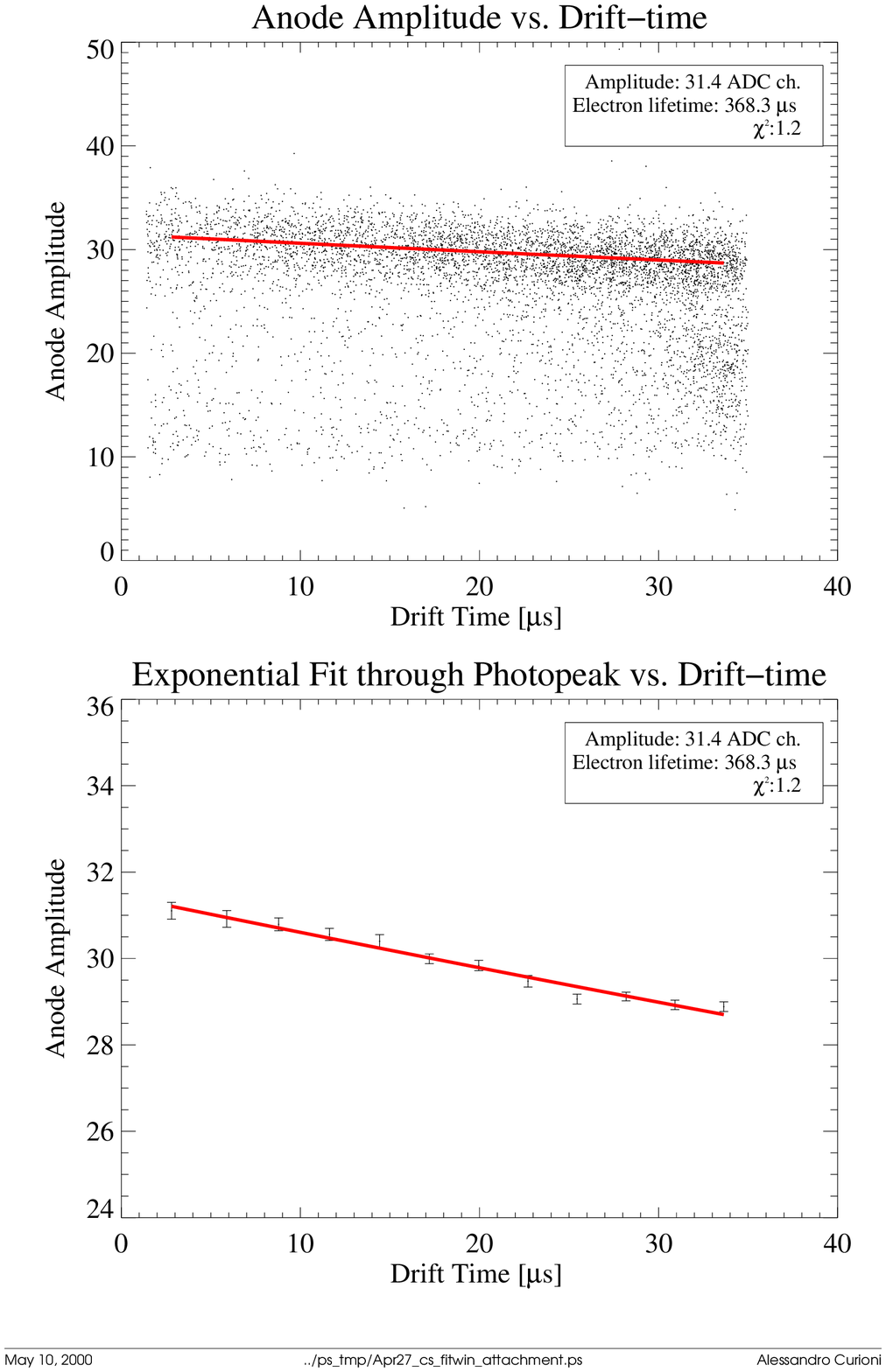,bbllx=77,bblly=420,bburx=580,bbury=727,width=.75\linewidth,clip=}
\caption{\label{fig_1} Anode amplitude (ADC ch.) {\it vs.} drift-time
($\mu$s). Drift-time is directly converted in $z$-position. The solid line is the result of an exponential
fit to the photopeak position for various slices in drift time.} 
\end{figure}

\begin{figure}[htb]
\centering
\epsfig{file=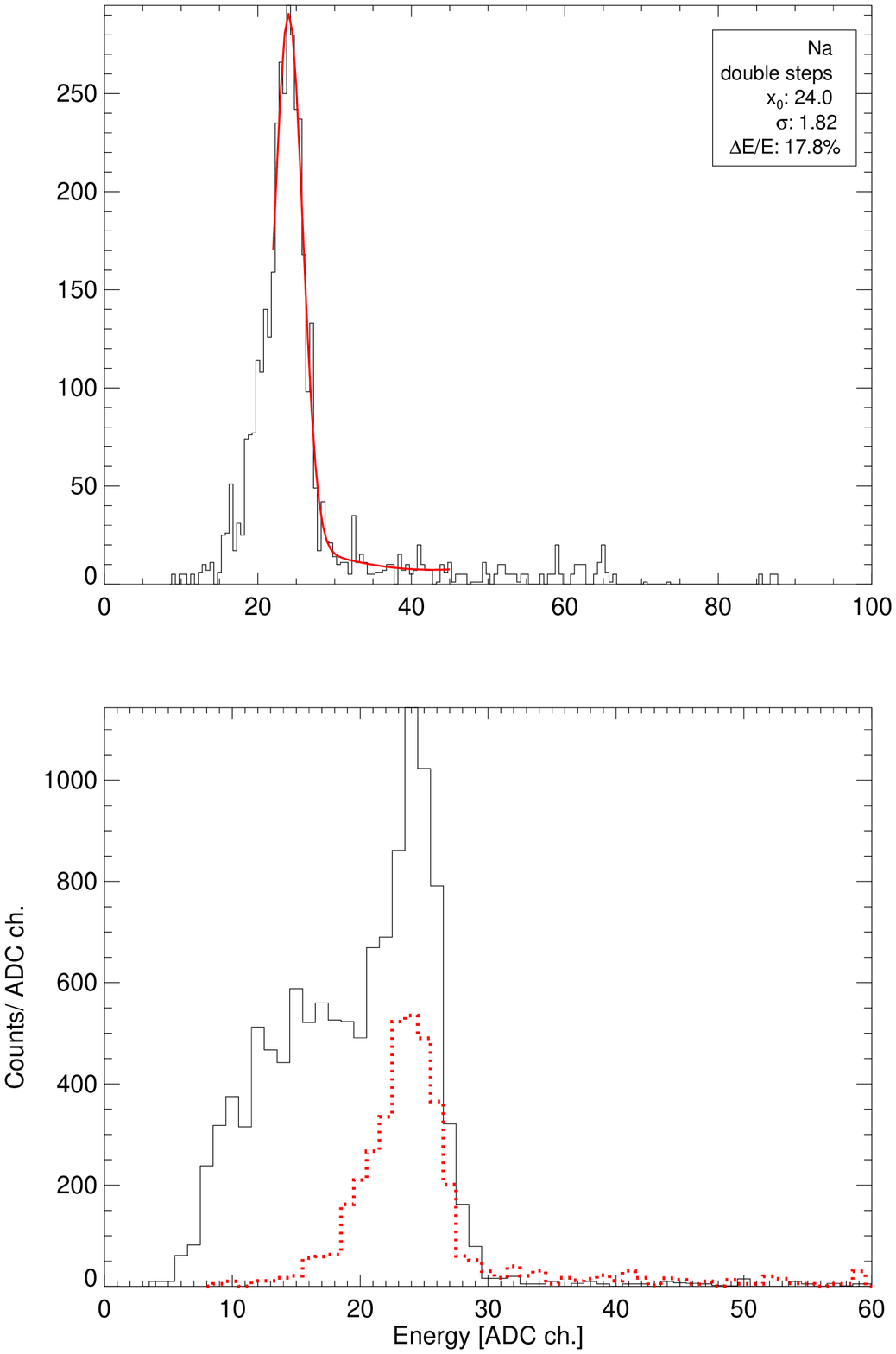,width=.7\linewidth,,bbllx=77,bblly=70,bburx=580,bbury=400,clip=}
\caption{\label{f:na_spec} \na\ 511~\keV \  energy spectrum for single interaction ({\it solid line}) and Compton interaction events ({\it dotted line}). Note the suppression of the Compton continuum and the large
enhancement of the peak-to-Compton ratio in the multiple interactions sample.} 
\end{figure}
\begin{figure}[htb]
\centering
\epsfig{file=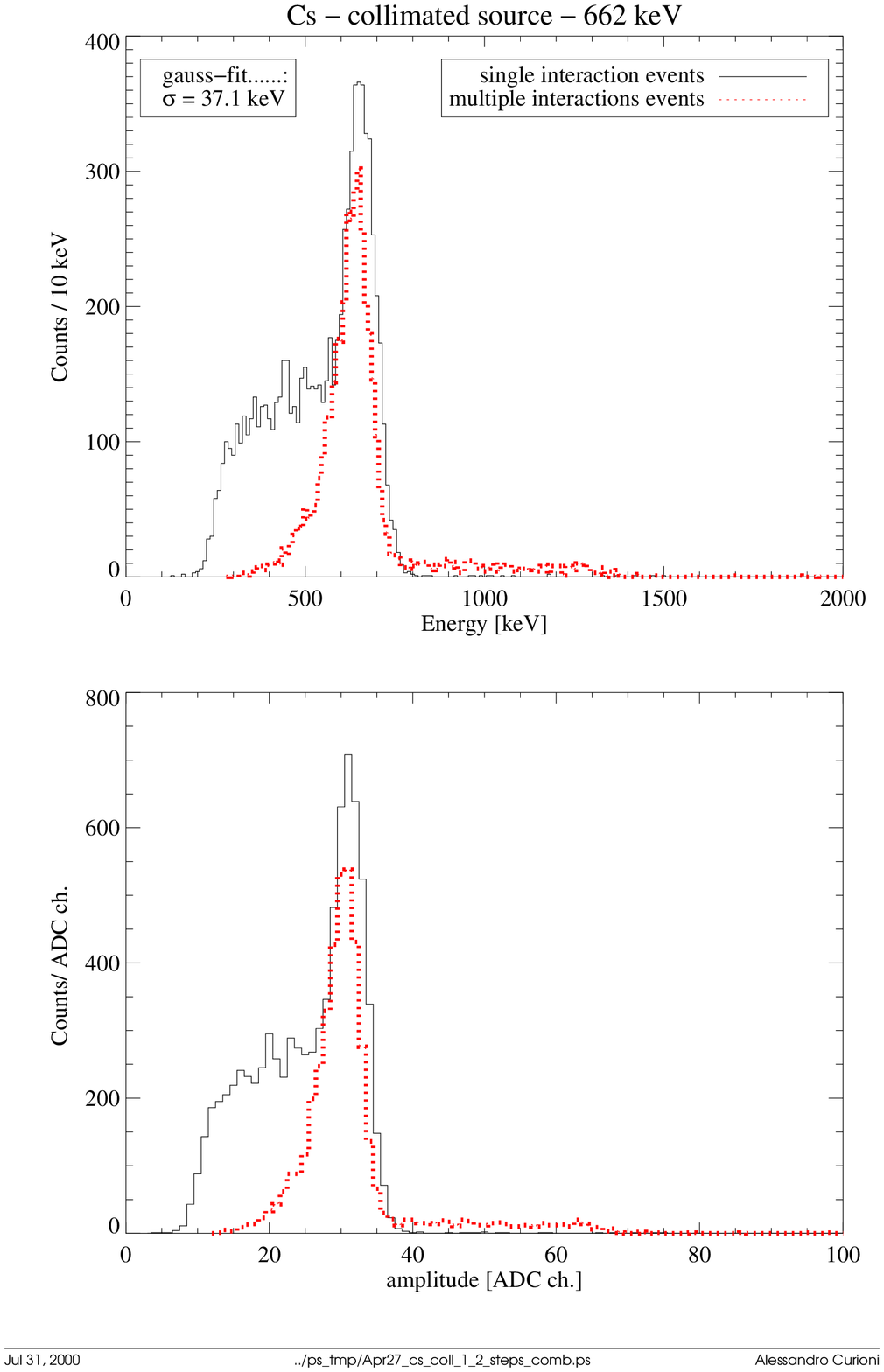,width=.7\linewidth,bbllx=77,bblly=70,bburx=580,bbury=400,clip=}
\caption{\label{f:cs_spec} \cs 662~\keV \ energy spectrum for single interaction ({\it solid line}) and Compton interaction events ({\it dotted line}). Note the suppression of the Compton continuum and the large
enhancement of the peak-to-Compton ratio in the multiple interactions sample.}
\end{figure}
\begin{figure}
\centering
\epsfig{file=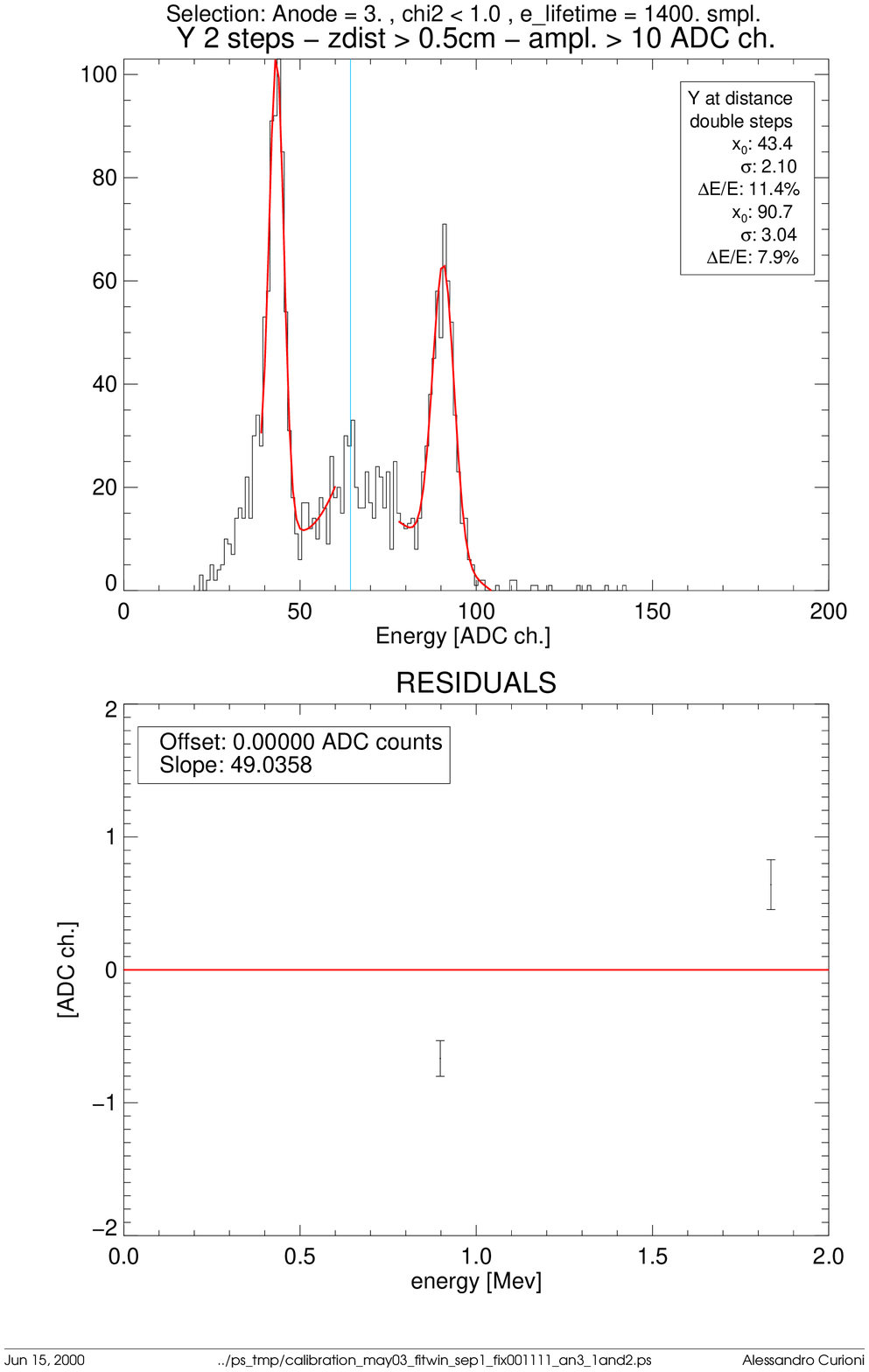,bbllx=77,bblly=420,bburx=580,bbury=733,
        width=.7\linewidth,clip=}
\caption{\label{fig_4} \yt\ 898~\keV \ and 1836~\keV \ energy spectrum for two interaction ("2--step") events. The source was placed about 2~m above the TPC.}
\end{figure}
\begin{figure}[htb]
\centering
\bigskip
\epsfig{file=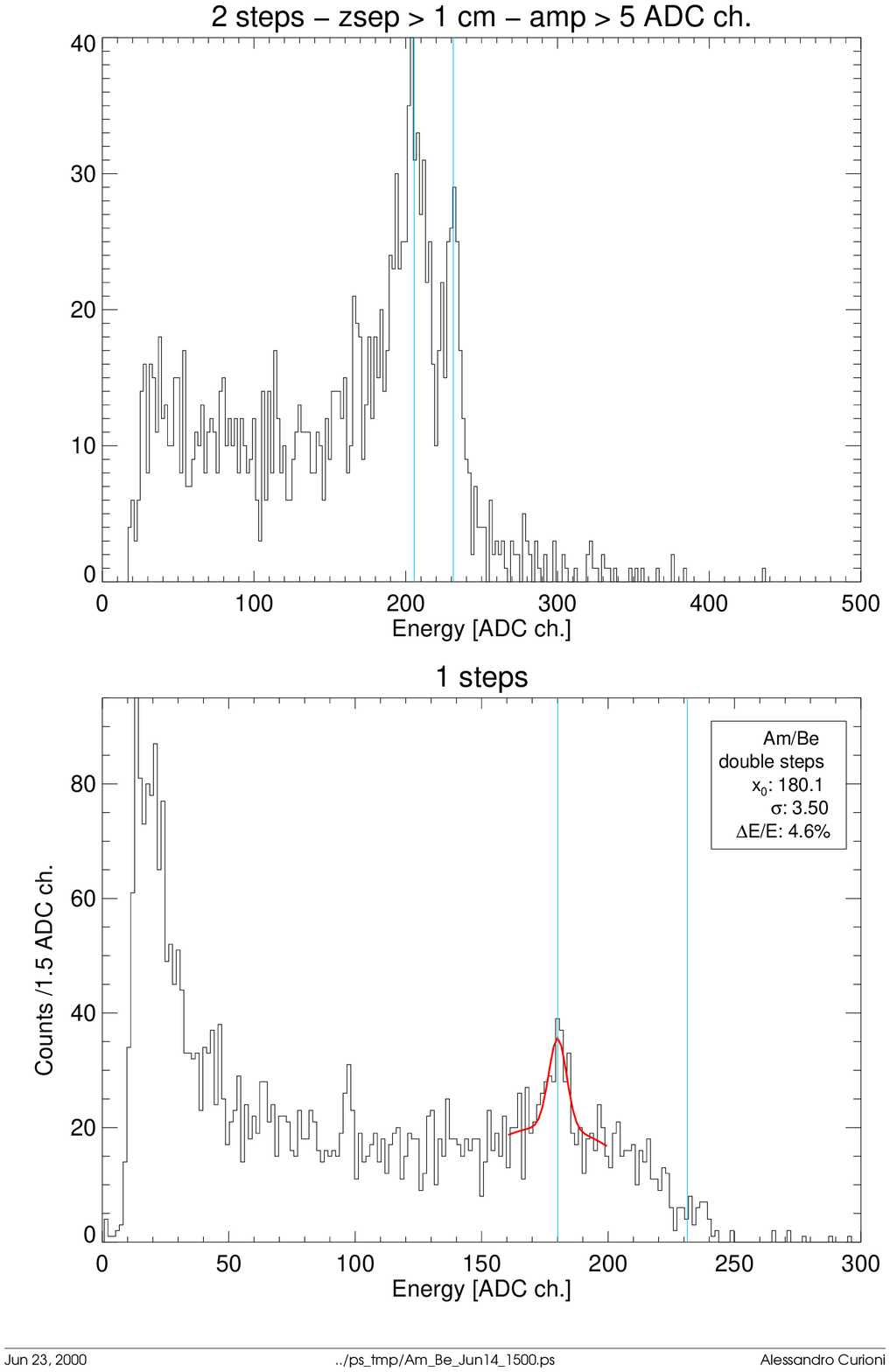,bbllx=77,bblly=410,bburx=580,bbury=728,
        width=.7\linewidth,clip=}
\caption{\label{fig_5} Am-Be energy spectrum for two interaction ("2--step")
events. The photopeak (4.43~\MeV) and the single escape peak (3.92~\MeV) are
clearly identified.}
\end{figure}

The response of the TPC to \g --rays was characterized in several experiments
with the electronics, trigger conditions and data acquisition system as used
during the May 99 flight. The following radioactive sources were used: \cs, \na\
and \yt\ with lines at 662~\keV, 511~\keV\ and 1275~\keV, 898~\keV\ and
1836~\keV, respectively. The sources were either collimated or placed at some
distance (2--4 meters) from the TPC. Data were also taken with an Am-Be neutron
source, which emits 4.43 \MeV\ gamma-rays, as well as with cosmic rays. \\ For
detectors with a large drift region like LXeGRIT, the Xe purity largely
determines the charge yield and thus the spectral response. To minimize the
charge loss due to attachment of free electrons to electro-negative impurities,
an efficient purification system was developed and has been fully discussed in
Xu 1998 \cite{r:fxu}.  The simple relation 
\be Q_{det}~=~Q_0 \cdot exp(-
\frac{z}{\lambda _{att}})
\label{eq_1}
\ee 
describes the detected charge $Q_{det}$, given an initial charge $Q_0$ at a
distance $z$ from the collection point. The parameter $\lambda _{att}$ is the
attenuation lenght for electrons drifting in liquid Xe. If not corrected, even a
modest (few percent) dependence of the collected charge on $z$ would deteriorate
the energy resolution. \\
The measurement of the absolute depth of an interaction in the light triggered
LXeTPC offers a direct method to infer the drifting electron lifetime and thus
the purity of the LXe.  Given a \g --ray line, in this case 662~\keV, the
procedure consists in measuring the photopeak amplitude for different $z$
(i.e. drift time) slices, as shown in Fig.~\ref{f:cs_slic}. The photopeak
dependence {\it vs.}  $z$ is then fitted with the relation \ref{eq_1} to yield
$\lambda _{att}$, i.e. the electron lifetime. The scatter plot of the amplitude
of the \cs\ events {\it vs.} drift time at 1~kV/cm is shown in Fig.~\ref{fig_1},
for single interaction events. The enhancement seen in correspondence to the
662~\keV\ full energy peak moves to lower pulse heights for longer drift time.
The measured drifting electron lifetime is about 360~$\mu$s: this lifetime
corresponds to an attenuation lenght of 80~cm, to be compared with the 7~cm
maximum drift lenght in our detector. A correction for attachment to impurities
is applied during event reconstruction on an event-by-event basis, removing the
dependence of the signal amplitude on the distance from the anode and so
significantly improving the spectral performance. \\
The spectral  performance of the LXeTPC, inferred from the analysis of either
single or multiple interaction events, is shown with the reconstructed energy
spectra of the \na, \cs, \yt\ and Am-Be sources (see Figs. \ref{f:na_spec},
\ref{f:cs_spec}, \ref{fig_4}, \ref{fig_5}). \\
Fig.~\ref{f:na_spec} shows the 511~\keV \ energy spectrum measured with a tagged
\na\ source: \na\ simultaneously emits a 1275~\keV\ \g --ray plus a positron,
promptly annihilating and producing two 511~\keV \ \g --rays in opposite
directions, so that one of the two 511~\keV \ photons can be used to tag the
other one, reducing 1275~\keV \ and pile-up events. In our case, when 511~\keV \
energy deposit was detected in a NaI(Tl) scintillation counter placed above the
TPC, a trigger was sent to the TPC. \\
Fig.~\ref{f:cs_spec} shows the energy spectrum of a collimated \cs\ source. In
both cases, we point out the large enhancement of the {\it
peak-to-Compton-ratio} obtained with multiple interactions events, compared to
the single interaction events spectrum. Fig.~\ref{fig_4} shows the \yt\ energy
spectrum obtained from events with two interactions ("2--step" events). In this
experiment the source was placed about 2~m above the TPC to simulate a parallel
\g-rays beam for imaging analysis, as discssed in the next section. Aside the
898 and 1836 \keV\ photopeaks, the single escape peak (1325 \keV) for 1836 \keV\
photons is detected.  Fig.~\ref{fig_5} shows the energy spectrum from an Am-Be
neutron source. Again "2--step" events were used. Since the cross-section for
pair-production is large at 4.4~\MeV, the single escape peak (3.92~\MeV) is
prominent.  Photopeak, single and double (detected in the ``1-step'' events
spectrum, not shown here) escape peaks and the Compton edge have been used in
the energy calibration. \\
\begin{figure}[htb]
\centering
\epsfig{file=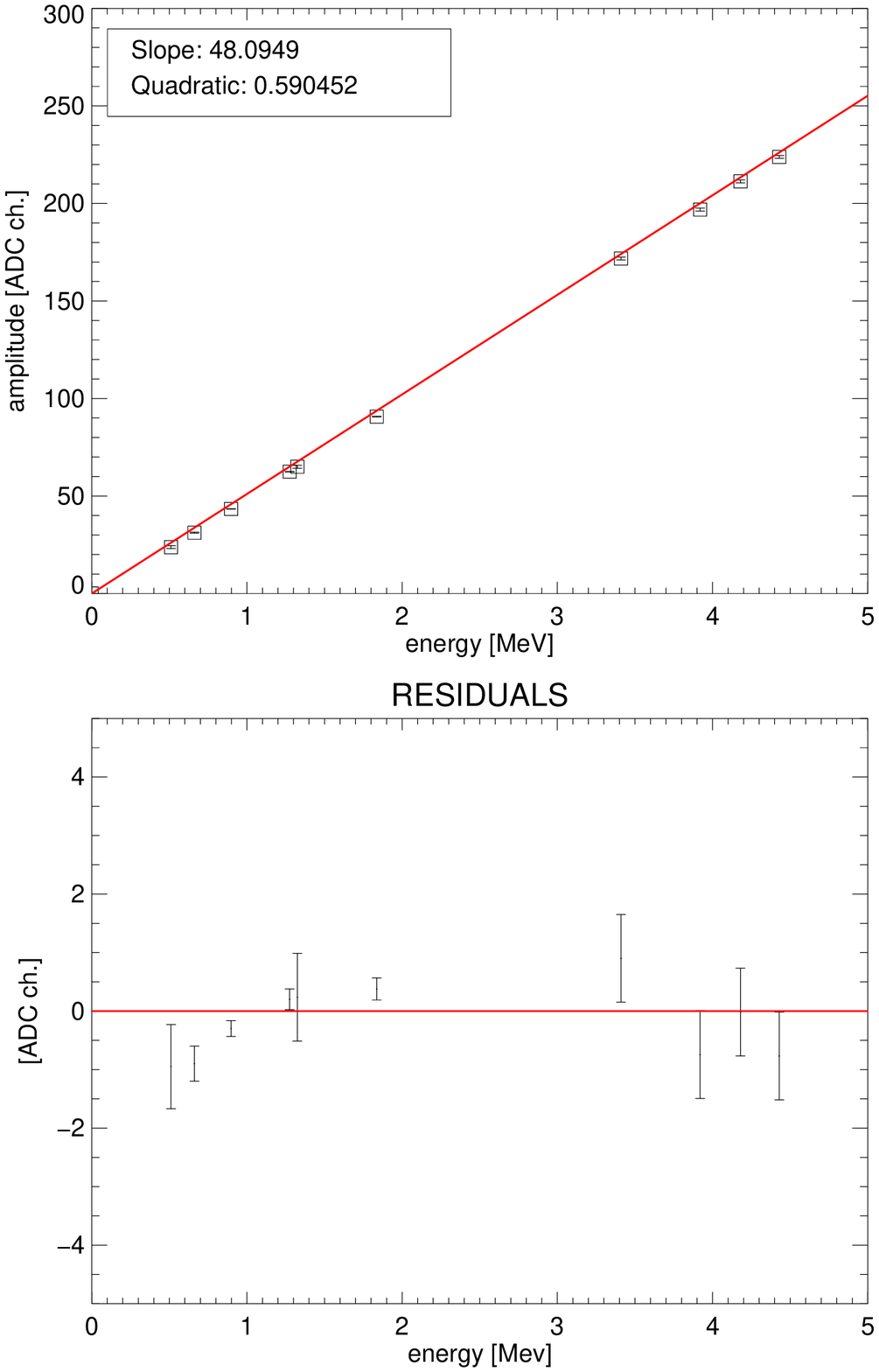,bbllx=77,bblly=410,bburx=580,bbury=735,width=.7\linewidth,clip=}
\caption{\label{fig_7} Energy calibration curve for one of the four LXeTPC anodes: anode amplitude {\it vs.}
energy fitted with $amp = b\cdot E + c\cdot E^2$. In the inlet, b = ``slope''
and c = ``quadratic''.}  
\end{figure}
\begin{figure}[htb]
\centering
\epsfig{file=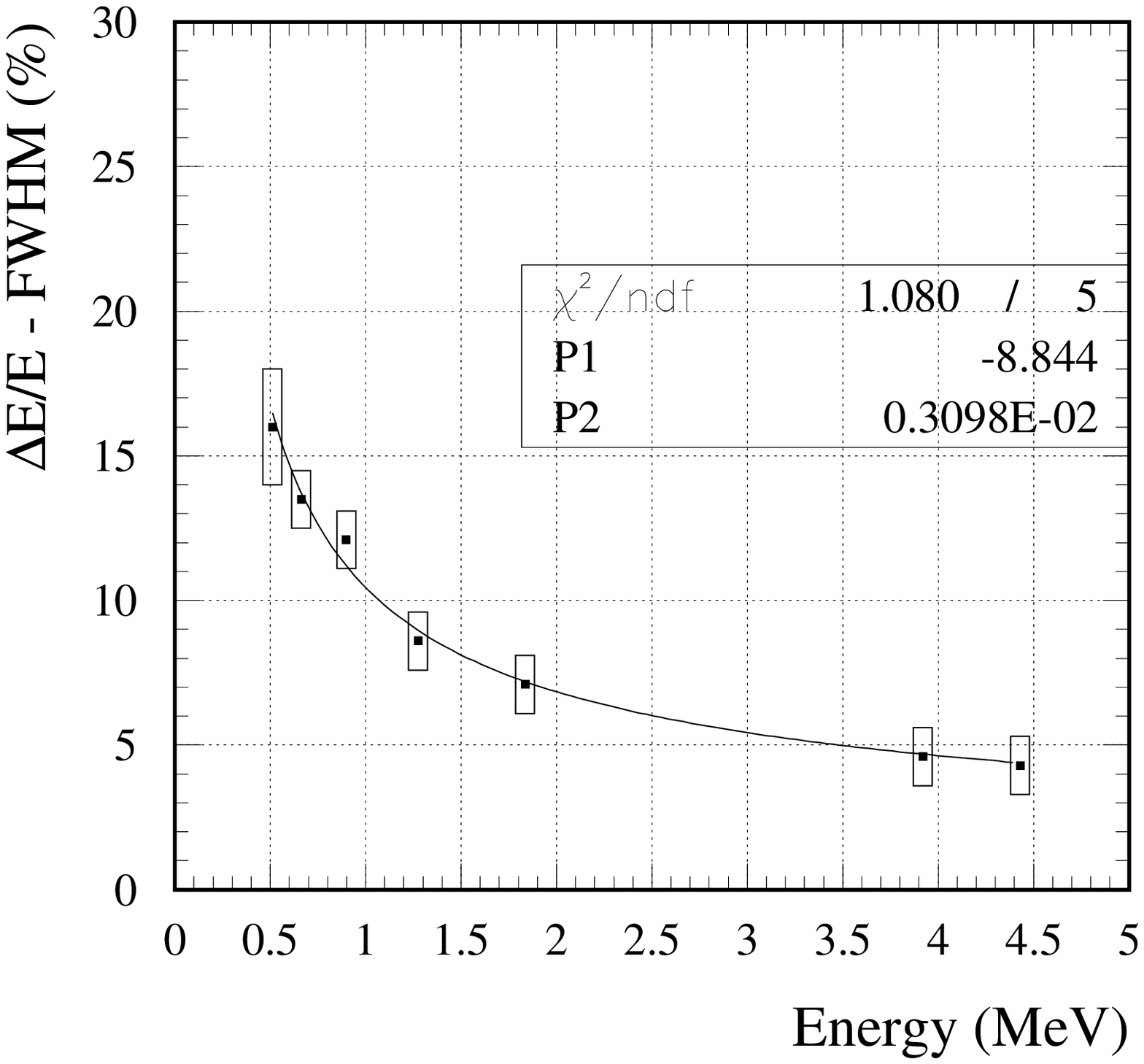,width=0.6\linewidth,clip=}
\caption{\label{fig_eres} Energy resolution {\it vs.} energy: $ \Delta
E_\mathrm{lxe}/E=8.8\% \: \sqrt{1 \MeV /E}$}
\end{figure}

The calibration curve obtained combining the \cs, \na,
\yt and Am-Be data is shown in Fig.~\ref{fig_7}. The non linear term is small,
indicating good proportionality over the energy range 511~\keV -- 4.43~\MeV.
The energy resolution of all the lines is consistent with a value of 
10~$\%$ (FWHM at 1~\MeV), scaling as $1/\sqrt{E}$. This
value includes electronics noise, shielding inefficiency and non linearities in
the gain response due to the electronics as well as the signal processing
analysis. The noise subtracted energy resolution is 8.8~$\%$ (FWHM at 1~\MeV)
(see Fig.~\ref{fig_eres}), in excellent agreement with expectations.
%
%
%

\subsection{Compton Imaging and Angular Resolution %
         \label{s:imaging}}
To test the Compton imaging performance of the telescope, the LXeTPC was
irradiated with a parallel beam of \g --rays from a \yt \ calibration source
placed at a distance of several meters. Here we present the results obtained
with \g --ray events, for 
which two interactions are recognized in the sensitive LXe volume
("2--step" events). Events with a single Compton scattering
followed by a photoabsorption are the simplest ones that can be used for 
imaging a source at a known location.  
The process of determining the imaging response breaks down into three parts:
\begin{figure}[htb]
\includegraphics[bb=50 300 570 760,width=.9\linewidth,clip]{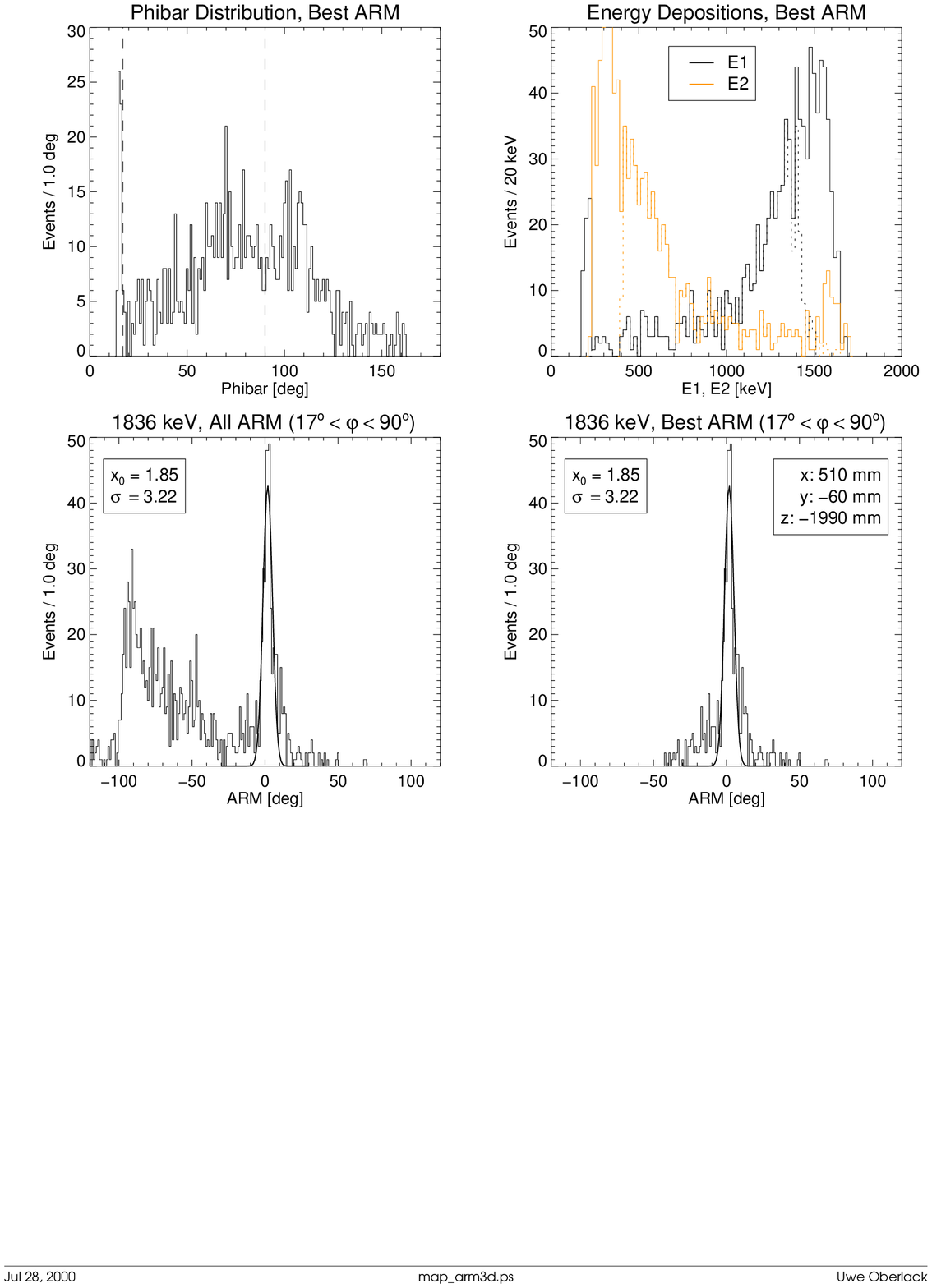}
\caption{\label{f:arm-two-steps}
{\it Top left:} \phibar \ distribution for two-step events in the full energy peak
of the 1836~\keV\ \yt \ line. {\it Top right:} energy deposit in each
interaction.  
{\it Bottom:} ARM spectra for two-step events in the full energy peak of the
1836~\keV\ \yt \ line.}
\end{figure}
\begin{enumerate}
\item Energy calibration, which is necessary for the relation of charge
  deposits to the electron mass $m_0 c^2$ in the calculation of the
  Compton scatter angle \phibar\ (equation~\ref{e:phibar:ij}). This includes
  correction for charge losses due to electron attachment according to the
  depth of the interaction (z-position), as described in section \ref{s:ecal}.
  The conversion from ADC counts to \keV \ is then based on the four anode
  calibration curves, such as that of Fig.~\ref{fig_7}. 

\item Determination of the time sequence of the two interactions. For
  events with two interactions,
  the LXeTPC provides no redundant measurement that would uniquely determine the
  interactions sequence. In the case of calibration data with known source
  position, the sequence with the minimum value of ``angular resolution measure'' (ARM) as defined below, can be chosen. For data with unknown source location, the decision on the correct sequence can be based on a maximum energy value
  assigned to the second step, since the photo-absorption cross section falls
  off steeply beyond 200~\keV\ in xenon (see
  \cite{SBoggs:00:comp_reconst,GJSchmid:99:tracking} for similar approaches).
  Alternatively, both sequences might be used in a likelihood analysis, which
  uses relative probabilities for each sequence with respect to a source
  position. Such extended instrumental response would be derived from Monte
  Carlo simulations. 
      
\item Derivation of the ARM  distribution,
  i.e., the distribution of scatter angles \phibar\ derived from Compton
  kinematics minus the geometrical scatter angles \phigeo\ between the known
  source position and the measured scatter direction, given by the two interaction
  locations:
  \begin{eqnarray}
  ARM &=& \phibar - \phigeo \\
  \cos \phibar_{ij}  
    &=& 1 - \frac{m_0 c^2}{E_{j}} + \frac{m_0 c^2}{E_{i}+E_{j}} \qquad
        i,j=\{1,2\} \textrm{ or } \{2,1\}
        \label{e:phibar:ij}	\\
  \cos\phigeo_{ij} &=& \frac{\vec{r}_{ij} \cdot \rsi}%
                       {|\vec{r}_{ij}| \: |\rsi|}
        \label{e:phigeo:ij}
  \end{eqnarray}
  $E_{i}$, $E_{j}$ denote the energy deposits presumed to be first and second,
  respectively. $\vec{r}_{ij} = \vec{r}_j - \vec{r}_i$ is the vector between the
  locations $i,j$ and \rsi\ is the vector between the source and location $i$.
\end{enumerate}

\begin{sloppypar}
\begin{wrapfigure}{R}{.55\textwidth}
\centering
\includegraphics[bb=263 76 492 362,width=.8\linewidth,angle=90,clip]{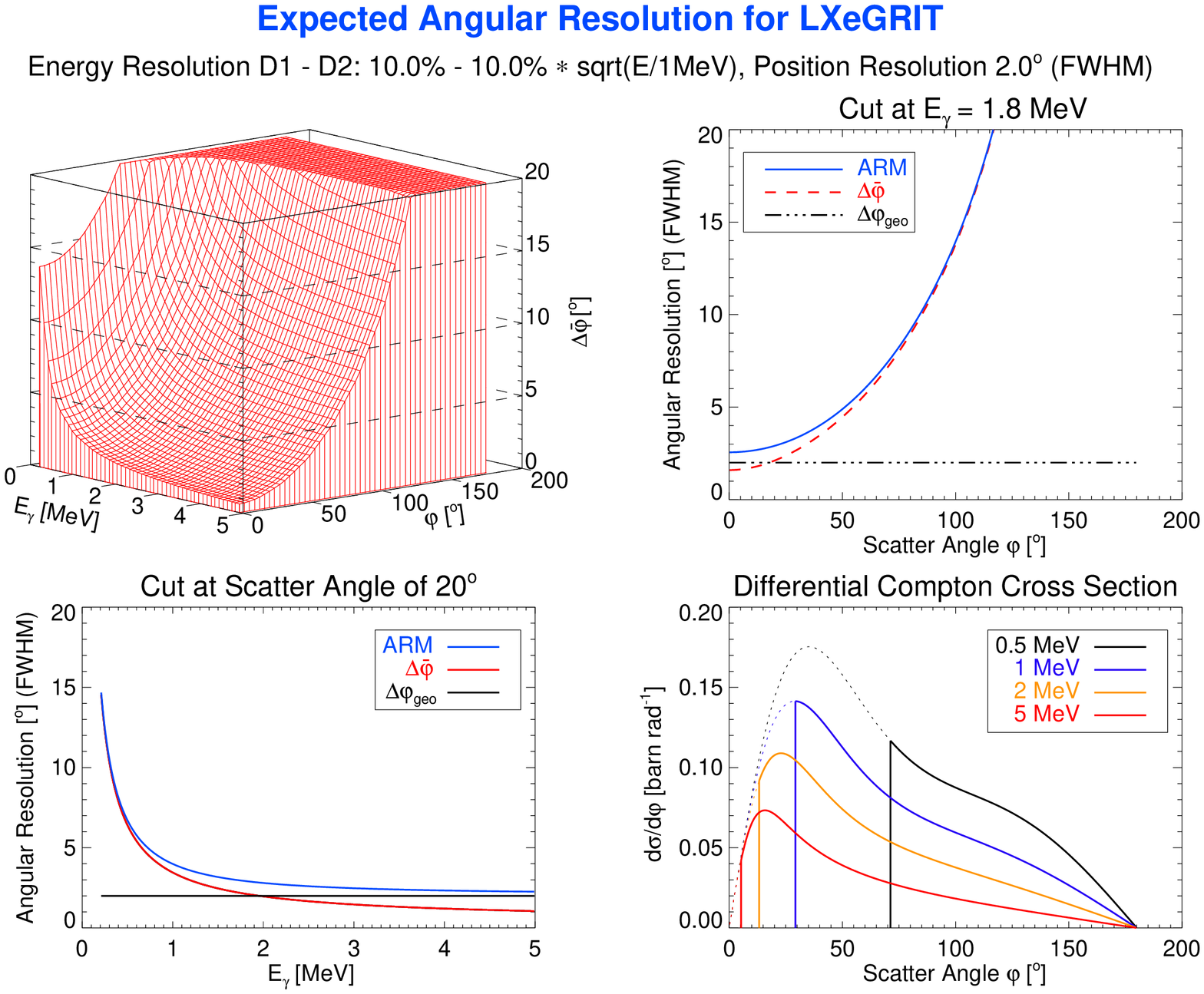}
\caption{\label{f:angres_vs_phi} Angular resolution (ARM) {\it vs.} \phibar \
for 1.8~\MeV \ photons (solid line) and contributions from energy resolution
(dashed line) and position resolution (dash-dotted line).}  
\end{wrapfigure}
For a \g-ray line source, as is the case for our calibration data, events can
be restricted to the full-energy peak, i.e., amplitudes $A_1 +A_2 \approx
A_\mathrm{line} \pm 2\,\sigma$. The \yt \  energy spectrum  with the two lines
at 898~\keV\ or 1836~\keV\ as obtained from two-step events was previously  shown in
Fig.~\ref{fig_4}  
Also, we require $\cos \phibar \ge
-1$ for a valid Compton sequence. Invalid sequences can result in  $\cos
\phibar$ values $<-1$. In fact, true Compton sequences with large scatter
angles, can also yield $\cos \phibar$ values $<-1$, due to the energy
uncertainty of the measurement.  
Back-scattered events would be rejected in any case since they provide a poor imaging response.
Results from the analysis of 1836~\keV\ ``2-step events'' from \yt \ at a
distance of about 2~m and at a zenith angle of $\sim 25\deg$ are shown in
Fig.~\ref{f:arm-two-steps}. The {\it Top left} 
panel shows the \phibar \ distribution for events in the full energy peak.  The
{\it Top right} panel shows the distribution of the energy deposit in each 
interaction. The {\it Bottom left} panel shows the  
ARM distribution  of the events in the full energy peak, derived without
selecting a time sequence, while the sequences with the smaller ARM value for a
given source position was selected for the {\it Bottom right} panel. The ARM
values for the two sequences 
are sufficiently separated, with the false sequence assuming a maximum around
$-93\deg$, such that choosing the smaller ARM, in the {\it Bottom right} figure,
does not bias the ARM peak for the correct sequence. The imaging response for
these two-step events consists of a narrow peak of $3.3\deg$ ($1\,\sigma$) for a
scatter angle selection of $17\deg < \phibar < 90\deg$. This angular resolution
value is consistent with expectations, as shown in Fig.~\ref{f:angres_vs_phi}:
for a scatter angle of $70\deg$ the resolution is about $7\deg$  (FWHM) at
1.8~\MeV\. 
The expected LXeGRIT angular resolution for a \g-ray line at 1.8~\MeV\ shown in
in this figure was calculated based on  the experimental energy resolution
(FWHM) of $ \Delta E/E=10\% \: \sqrt{1\MeV /E}$ and assuming a 2\deg \
geometrical angle contribution due to the uncertainty in the position
measurement. 

The non-Gaussian wings in the measured ARM distribution may be due to
three-interaction events that are seen as two steps within 
the resolution of the detector, or to imperfections in the off-line fitting of
the anode waveform. 
Note that with a minimum spatial separation of $\sim 2$~cm between the two
interactions, the 
angular response is fully determined by the energy resolution and by the lowest
energy threshold, which in this data set was about 170~\keV. This imposes an
energy-dependent lower threshold in \phibar. 
For events with three or more interactions, the correct sequence of scatterings
can be inferred from the redundant position and energy information measured by
the TPC for each interaction. The performance of a Compton sequence
reconstruction algorithm on  LXeGRIT  ``3-step'' events is reported in
Oberlack et al. 2000 \cite{oberlack_2}. In this paper we show that the ARM\
distribution for the 1836~\keV\ line of \yt \ events has a width of $\sim 3\deg$
($1\,\sigma$), consistent with the result obtained with the ``2-step'' events.  
\end{sloppypar}

\section{Conclusions}

During 1999 the spectral and imaging performance of the LXeGRIT instrument was
extensively measured with a variety of  \g --ray sources, spanning the energy
range 511~\keV --4.4~\MeV, in preparation for the May~'99 balloon flight.  
Results can be summarized as follows:  
\bi
\item the LXeTPC energy response shows a good linearity over the measured.
      energy range 
\item the LXeTPC energy resolution, at an electric field of 1kV/cm, is $8.8\%$ FWHM at 1~\MeV\, scaling with $1/\sqrt{E}$.
      This value is consistent with previous results, obtained however with  simple gridded ionization chambers with drift regions of a few mm to a few cm (see
      for example Aprile et al. 1991 \cite{EAprile:91:performance}).    
\item the 3D imaging capability of the TPC 
      has been confirmed as a powerful tool to suppress background events and
      to enhance the {\it peak-to-Compton ratio} in the measured spectral distributions.    
\item the TPC works as a Compton telescope: \g --ray sources can be imaged
      with an angular resolution consistent with expectations (3 \deg \
      RMS for 1.8~\MeV \ \g --rays).
\item the overall performance of the TPC in the laboratory shows excellent stability in time,
      from the liquid purity to the high voltage, from the cryogenics to the analog and 	digital electronics.	 
\ei

These results confirm the applicability of this new detector concept to \MeV \
\g --ray astrophysics.  

\acknowledgments  

This work was supported by NASA under grant NAG5-5108. 

\small
\bibliography{calib_SPIE00}  
\bibliographystyle{spiebib}   

\end{document}